\documentclass[aps,prl,reprint,superscriptaddress,nofootinbib]{revtex4-1}

\usepackage{graphicx}
\usepackage{amsmath}
\usepackage{amssymb}
\usepackage{amsfonts}
\usepackage{verbatim}
\usepackage{mathrsfs}
\usepackage{array}
\usepackage{layout}
\usepackage{textcomp}
\usepackage{latexsym}
\usepackage{slashed}
\usepackage{rotating}
\usepackage{hyperref}
\usepackage{booktabs}
\usepackage{relsize}
\usepackage{multirow}
\usepackage{orcidlink}

\newcommand{\Eq}{&=&}

\newcommand{\hs}[1]{{\hspace{#1}}}
\newcommand{\vs}[1]{{\vspace{#1}}}

\newcommand{\mb}[1]{{\mathbf{#1}^{}}}
\newcommand{\tx}[1]{{\text{#1}^{}}}
\newcommand{\nn}{\nonumber\\}
\newcommand{\rx}[2]{{\raisebox{#1}{#2}}}

\newcommand{\dd}{{\mathrm{d}}}

\newcommand{\UL}{U$(1)^{}_\tx{L\,}$}
\newcommand{\vh}{\upsilon_h}
\newcommand{\vphi}{\upsilon_\phi}

\begin{document}

\title{TeV-scale Unification of Light Dark Matter and Neutrino Mass}

\author{Cheng-Wei Chiang\orcidlink{0000-0003-1716-0169}}
\email{chengwei@phys.ntu.edu.tw}
\affiliation{Department of Physics, National Taiwan University, Taipei 10617, Taiwan}
\affiliation{Physics Division, National Center for Theoretical Sciences, National Taiwan University, Taipei 106319, Taiwan}

\author{Shu-Yu Ho\orcidlink{0000-0003-0826-2956}}
\email{shuyuho@as.edu.tw}
\affiliation{Institute of Physics, Academia Sinica, Nangang, Taipei 11529, Taiwan}

\author{Van Que Tran\orcidlink{0000-0003-4643-4050}}
\email{vqtran@phys.ncts.ntu.edu.tw}
\affiliation{Physics Division, National Center for Theoretical Sciences, National Taiwan University, Taipei 106319, Taiwan}
\affiliation{Phenikaa Institute for Advanced Study, Phenikaa University, Nguyen Trac, Duong Noi, Hanoi 100000, Vietnam}


\begin{abstract}
We demonstrate that TeV-scale heavy neutral leptons (HNLs) responsible for inverse-seesaw neutrino mass generation can simultaneously fix the cosmological abundance and decay properties of dark matter (DM).\,\,The spontaneous breaking of lepton number gives rise to a pseudo-Nambu-Goldstone boson that serves as a light DM candidate, whose mass originates from a small explicit symmetry-breaking term.\,\,The same HNLs that generate neutrino masses produce the DM via freeze-in and mediate its decay into neutrinos, leading to a tight correlation among neutrino masses, DM relic abundance, and DM lifetime.\,\,For collider-accessible TeV-scale HNLs, the observed relic density and lifetime constraints point to sub-GeV DM, yielding observable neutrino signals at JUNO and next-generation detectors such as Hyper-Kamiokande and DUNE.\,\,This framework establishes a predictive and experimentally testable link between neutrino mass generation and DM.
\end{abstract}

\maketitle

\textit{Introduction.}\textemdash The origin of neutrino mass and the identity of dark matter (DM) remain two of the most profound open questions in particle physics and cosmology.\,\,Neutrino oscillation experiments have established that neutrinos are massive, providing the first direct evidence of physics beyond the Standard Model (SM), yet the mechanism responsible for their tiny masses is still unclear~\cite{Super-Kamiokande:1998kpq,SNO:2001kpb}.\,\,At the same time, cosmological and astrophysical observations point to a non-baryonic DM component whose particle nature remains elusive~\cite{Planck:2018vyg}.\,\,A unified and experimentally testable framework linking neutrino mass generation and DM is therefore highly desirable.

Heavy neutral leptons (HNLs) can provide a compelling explanation for neutrino mass through the seesaw mechanism~\cite{Minkowski:1977sc,Yanagida:1979as,Gell-Mann:1979vob,Mohapatra:1979ia,Schechter:1980gr,Cheng:1980qt,Schechter:1981cv,
Foot:1988aq}, and may also account for the baryon asymmetry of the Universe via leptogenesis~\cite{Fukugita:1986hr,Akhmedov:1998qx,Asaka:2005pn,Davidson:2008bu,
Hambye:2016sby}.\,\,In particular, the inverse seesaw mechanism naturally generates sub-eV neutrino masses while allowing the HNLs to reside at the TeV scale~\cite{Mohapatra:1986bd,Mohapatra:1986aw,Dias:2012xp}.\,\,Such TeV-scale states are actively searched for in collider experiments through multi-lepton final states and electroweak production channels~\cite{CMS:2015qur,ATLAS:2015gtp,CMS:2016aro,CMS:2018iaf,
ATLAS:2019kpx,CMS:2024xdq}.\,\,For smaller active-sterile mixing angles and lighter masses, they may yield displaced-vertex signatures, motivating dedicated long-lived particle searches~\cite{CMS:2022fut,ATLAS:2022atq,CMS:2024hik,ATLAS:2025uah}.\,\,Despite this rich phenomenology, conventional inverse-seesaw constructions neither predict a viable DM candidate nor correlate neutrino mass parameters with cosmological observables~\cite{Abada:2014zra}.

Pseudo-Nambu-Goldstone bosons (pNGBs) arising from spontaneous lepton-number breaking, commonly referred to as Majorons, have long been investigated as DM candidates~\cite{Chikashige:1980qk,Pilaftsis:1993af}.\,\,Their cosmological production and decay into neutrinos have been explored in various contexts~\cite{Frigerio:2011in, Abe:2020ldj, Cheng:2020rla, Mohapatra:2021ozu, Manna:2022gwn, King:2024idj, Bernal:2025qkj}.\,\,In most existing constructions, however, the parameters governing neutrino mass generation, DM relic abundance, and DM lifetime are treated as completely independent inputs, and a direct connection to collider-accessible neutrino mass models is missing.

\begin{figure}[b!]
\begin{center}
\includegraphics[width=0.46\textwidth]{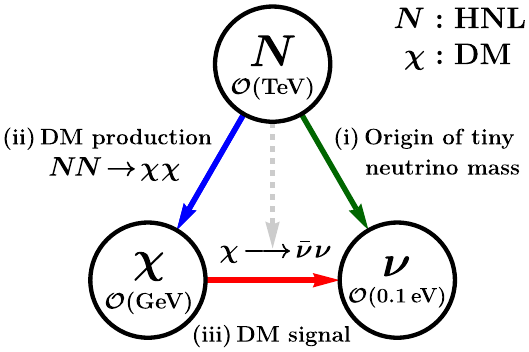}
\caption{Triangular connection for sub-GeV DM production, DM signal, and neutrino mass generation via TeV-scale HNLs.
\label{fig:model}}
\end{center}
\end{figure}

In this Letter, we demonstrate that a minimal extension of the inverse seesaw model can simultaneously determine neutrino mass, DM production, and DM decay in a coherent framework and thus in a predictive manner.\,\,The spontaneous breaking of lepton number introduces a pNGB that serves as a light DM candidate, whose mass arises from a small explicit breaking term~\cite{Barger:2008jx,Bodas:2020yho,Kim:2024tnn,Biswas:2024wbz}.\,\,Remarkably, the same TeV-scale HNLs responsible for neutrino mass generate DM through freeze-in and mediate its decay into neutrinos.\,\,Consequently, neutrino masses, DM relic abundance, and DM lifetime are determined by the same underlying parameters and cannot be chosen independently.

This model structure is seen to have tight and predictive connections among three sectors, with the HNLs playing a pivotal role, as illustrated schematically by the triangular diagram in Fig.\,\ref{fig:model}\,:\,(i) inverse-seesaw neutrino mass generation, (ii) freeze-in production of DM from TeV-scale HNLs, and (iii) DM decay into neutrinos.\,\,The lepton-number symmetry-breaking scale that controls the HNL masses simultaneously suppresses the DM decays, ensuring its cosmological longevity while permitting observable neutrino signals.

For collider-accessible TeV-scale HNLs, imposing the observed relic abundance and current lifetime bounds restricts the DM mass to the sub-GeV regime.\,\,In this range, the DM decay into neutrinos is testable at JUNO~\cite{Akita:2022lit} and next-generation detectors such as Hyper-Kamiokande~\cite{Bell:2020rkw} and DUNE~\cite{Arguelles:2019ouk}.\,\,The discovery of TeV-scale HNLs would therefore imply a correlated neutrino signal from the DM decay, establishing a concrete and experimentally accessible bridge among collider physics, cosmology, and neutrino astronomy.

\textit{Model.}\textemdash We consider a minimal extension of the SM by introducing singlet fermions $N_R^{}$ and $S_L^{}$, along with a complex singlet scalar $\phi$ carrying lepton number under a global \UL symmetry.\,\,The relevant particle content and their quantum numbers are summarized in Tab.\,\ref{tab:1}.\,\,In particular, the lepton number assignments of the new fields are chosen such that the interactions $\phi\, N_R^{} N_R^{}$ and $\phi\, S_L^{} S_L^{}$ are forbidden for simplicity.\,\,The relevant interactions are
\begin{eqnarray}
\mathcal{L} 
\supset\,
- \,\overline{E_L^{}} \widetilde{H} \, {\cal Y}_D^{} N_R^{} 
- \overline{S_L^{}} \, {\cal Y}_N^{} N_R^{} \, \phi
- \!\! \frac{1}{2} \, \overline{S_L^{}} \, \mu_S^{} S_L^c
+ \text{H.c.}
\,,
\label{eq:Yukawa}
\end{eqnarray}
where $E_{L}^{} = (\,\hat{\nu}_{L}^{}~~\ell_{L}^-\,){}^\tx{T}$ is the SM left-handed doublet, $\widetilde{H} = i \, \sigma^2 H^\ast$ with $H$ the SM Higgs doublet and $\sigma^2$ the second Pauli matrix.\,\,Here ${\cal Y}_D^{}$ and ${\cal Y}_N^{}$ are $3 \times 3$ Yukawa matrices, while $\mu_S^{}$ is a small $3 \times 3$ Majorana mass matrix that softly breaks lepton number and realizes the inverse seesaw mechanism~\cite{Mohapatra:1986bd,Mohapatra:1986aw,Dias:2012xp}.

\begin{table}
\caption{Quantum numbers of the relevant SM and new particles in the model, where $e_{Rj}^{}$, $E_{Lj}^{} = (\,\hat{\nu}_{Lj}^{}~~\ell_{Lj}^-\,){}^\tx{T}$, and $H$ are the SM right-handed singlet, left-handed doublet, and Higgs doublet, respectively, with $j = 1, 2, 3$ being the generation index.}
\begin{ruledtabular}
\begin{tabular}{c|ccc|ccc}
& ~$e_{Rj}^{}$~ & ~$E_{Lj}^{}$~ & ~$H$~ & ~$N_{Rj}^{}$~ & ~$S_{Lj}^{}$~ & ~$\phi$~  
\\[0.05cm]\hline
~\,SU$(2)_{\!L}^{}\vphantom{|_|^|}$~         
& ~$\mb{1}$~ & ~$\mb{2}$~ & ~$\mb{2}$~ & ~$\mb{1}$~ & ~$\mb{1}$~ & ~$\mb{1}$~  
\\[0.05cm]
~\,U$(1)_\tx{Y}^{}$~ & ~$-1$~ & ~$-1/2$~ & ~$1/2$~ & ~$0$~ & ~$0$~ & ~$0$~ 
\\[0.05cm]
~\,U$(1)_\tx{L}^{}$~ & ~$1$~ & ~$1$~ & ~$0$~ & ~$1$~ & ~$2$~ & ~$1$~ 
\\[0.05cm]
~$\tx{spin}$~ & ~$1/2$~ & ~$1/2$~ & ~$0$~ & ~$1/2$~ & ~$1/2$~ & ~$0$~ 
\end{tabular}
\label{tab:1}
\end{ruledtabular}
\end{table}

After spontaneous symmetry breaking, the scalar fields are expanded around their vacuum expectation values as
\begin{eqnarray}
H 
\,=\,
\begin{pmatrix}
0 \\
\dfrac{1}{\sqrt{2}} (\vh^{} + h)
\end{pmatrix}
~,\quad
\phi 
\,=\,
\frac{1}{\sqrt{2}} (\vphi^{} + \rho + i\chi)
~,
\label{eq:LR}
\end{eqnarray}
where $\vh^{} \simeq 246\,\,\text{GeV}$ and $\vphi^{}$ correspond to electroweak and lepton-number breaking scales, respectively.\,\,The fields $h$ and $\rho$ are CP-even scalars,\footnote{We assume negligible mixing between them, such that $h$ corresponds to the observed Higgs boson, while $\rho$ is a heavier scalar state.} while $\chi$ is a pNGB associated with spontaneous lepton-number breaking.

Working in the one-generation limit of the leptons and assuming $\mu_S^{} \ll m_D^{} \ll m_N^{}$, with $m_D^{} = {\cal Y}_D^{} \vphi^{}/\sqrt{2}$ and $m_N^{} = {\cal Y}_N^{} \vphi^{}/\sqrt{2}$, the mass eigenstates of neutral leptons can be obtained perturbatively.\,\,In particular, the light neutrino state is approximately
\begin{eqnarray}
\label{eq:nu}
\nu 
\,\simeq\, 
\big(1 - \xi^2/2\big) \,\hat{\nu}_L^{} - \xi\, S_L^{} + \xi \,\, \omega\, N_R^c
~,
\end{eqnarray}
where $\xi \equiv m_D^{}/m_N^{}$ and $\omega \equiv \mu_S^{}/m_N^{}$,
while the HNL states $N_{1,2}^{} \sim N_R^c \mp S_L^{}$ are nearly degenerate and form a quasi-Dirac lepton with a small admixture of the left-handed neutrino, $\hat{\nu}_L^{}$, of order $\xi$.\,\,The light neutrino mass is given by
\begin{eqnarray}
m_\nu^{}
\,\simeq\,
0.1\,\,\text{eV}
\bigg(\frac{\mu_S^{}}{1\,\,\text{keV}}\bigg)
\bigg(\frac{m_D^{}}{10\,\,\text{GeV}}\bigg)^{\hs{-0.12cm}2}
\bigg(\frac{m_N^{}}{1\,\,\text{TeV}}\bigg)^{\hs{-0.12cm}-2}
~.
\label{eq:mnu}
\end{eqnarray}
This illustrates a key advantage of the inverse seesaw\,:\,sub-eV neutrino masses can be realized with TeV-scale heavy states, in contrast to the canonical seesaw which typically requires $m_N^{} \sim 10^{14}\,\,\tx{GeV}$.\,\,The smallness of $\mu_S$ softly breaks lepton number, which is restored in the limit $\mu_S^{} \to 0$.

To generate a mass for the pNGB, we introduce a linear soft-breaking term in the scalar potential~\cite{Barger:2008jx,Bodas:2020yho,Kim:2024tnn},
\begin{eqnarray}
{\cal V}_{\rm soft} 
\,=\, 
- \, \frac{1}{\sqrt{2}} \, \kappa_\phi^3 \, \text{Re}(\phi)
~,
\end{eqnarray}
which yields $m_\chi^2 = \kappa_\phi^3/(2 \, \vphi^{})$ with $\kappa_\phi^{} > 0$. 
The pNGB nature of $\chi$, characterized by $m_\chi \ll \vphi^{}$, therefore requires $\kappa_\phi^{} \ll \vphi^{}$~\cite{Coito:2021fgo}.\,\,An additional virtue of this linear soft-breaking term is that it can evade the cosmic domain wall problem~\cite{Zeldovich:1974uw}, as pointed out in Ref.\,\cite{Barger:2008jx}.

An accidental $\mathbb{Z}_2$ symmetry (not a subgroup of \UL) can be identified in the scalar sector, under which the real fields transform as $\rho \to \rho$ and $\chi \to -\chi$.\,\,This symmetry can be understood as a remnant of a charge-conjugation transformation, $\phi \to \phi^\ast$, acting on the complex scalar.\,\,Consequently, the pNGB $\chi$, being $\mathbb{Z}_2$-odd, is stable at the renormalizable level and provides a natural DM candidate.\,\,However, the Yukawa interaction ${\cal Y}_N^{}$ in Eq.\,\eqref{eq:Yukawa} explicitly breaks this symmetry, rendering $\chi$ metastable.\,\,Its dominant decay channel is $\chi \to \bar{\nu}\nu$, with a rate strongly suppressed by the small neutrino masses and the large scale $\vphi^{}$.\,\,As we show below, $\chi$ remains cosmologically long-lived and can account for the observed DM abundance.\,\,This decay channel also provides a distinctive phenomenological signature at neutrino detectors.\,\,Details of the scalar potential, mass spectrum, and interactions in the model are provided in the Supplemental Material.

\textit{Dark Matter Production and Decay.}\textemdash To achieve the observed DM relic abundance, we employ the freeze-in mechanism, where DM was produced non-thermally from the thermal plasma in the early Universe~\cite{Hall:2009bx}.\,\,In particular, we consider the infrared freeze-in, where the reheating temperature of the Universe, $T_R^{}$, is much larger than the mass of the new CP-even Higgs boson, and the final abundance of DM is insensitive to $T_R^{}$.

For sufficiently large Yukawa coupling ${\cal Y}_D^{}$ and $m_N^{} < T_R^{}$, the HNLs thermalize with the SM plasma~\cite{Manna:2022gwn,King:2024idj,Barman:2022scg} (see the Supplemental Material with the detailed discussion), while the DM candidate $\chi$, due to its feeble interactions, never attains thermal equilibrium and is instead produced via freeze-in.\,\,DM production can in principle proceed through both SM particles and HNLs.\,\,In this Letter, we focus on the regime where the portal coupling between the CP-even scalars is negligible (which can be realized by an enhanced Poincar\'e symmetry~\cite{Foot:2013hna}), such that DM is predominantly generated through HNL annihilation.\footnote{Non-thermal production of pNGB DM from SM particles, arising from sizable portal couplings (i.e., scalar mixing), has been studied in Refs.\,\cite{Abe:2020ldj,Bernal:2025qkj}.}\,\,The relic abundance is then governed by the Yukawa interactions and HNL masses, and is largely insensitive to the scalar portal.\,\,This setup also suppresses direct detection signals, as the coupling between DM and SM fields is highly suppressed.

\begin{figure}[t!]
\begin{center}
\includegraphics[width=0.48\textwidth]{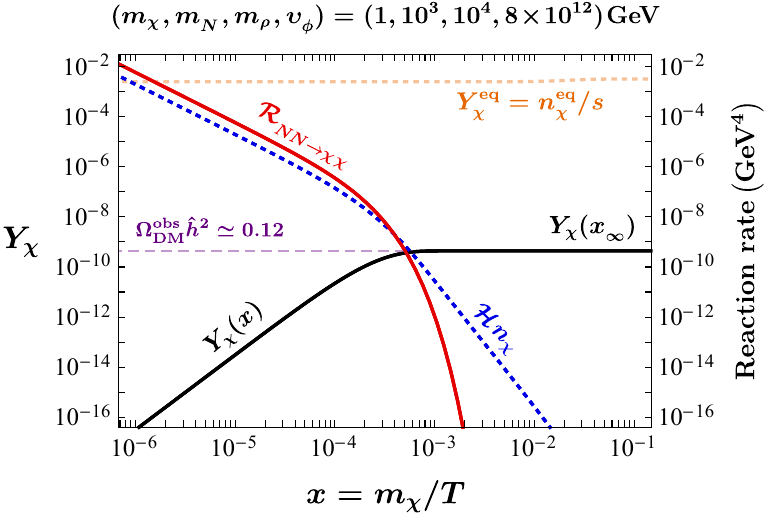}
\vs{-0.5cm}
\caption{Time evolution of the comoving DM number density, reaction rate, and Hubble expansion rate.}
\label{fig:Ychivsx}
\end{center}
\end{figure}

We further assume a hierarchical mass spectrum $m_\rho^{} \gg m_N^{} \gg m_\chi^{}$.\,\,In this regime, the Boltzmann equation governing the evolution of the DM number density $n_\chi(t)$ simplifies to
\begin{eqnarray}
\frac{d \, n_\chi}{d \, t}
+
3 \, {\cal H} \, n_\chi
\,=\,
{\cal R}_{N \! N \to \chi \, \chi}^{}
~,
\label{Boltzmann}
\end{eqnarray}
where ${\cal H}$ is the Hubble parameter during the radiation-dominated era, and ${\cal R}_{N \! N \to \chi \, \chi}$ denotes the total production rate from HNL annihilation.\,\,The annihilation processes are dominated by the $s$-channel exchange of $\rho$.\,\,The corresponding reaction rate is given by~\cite{Gondolo:1990dk}
\begin{eqnarray}
\hs{-0.6cm}
{\cal R}_{N \! N \to \chi \, \chi}^{}
\Eq
2\,\big(n_N^\tx{eq}\big)^{\!\!2} 
\sum_{k = 1,2}
\big\langle \sigma_{N_{k\tx{M}}^{} N_{k\tx{M}}^{} \to \chi \, \chi}^{} v_\tx{rel}^{} \big\rangle
\nn
&\simeq&
\frac{\lambda_\phi^2 \, m_N^2 \, T^2}{128\,\pi^5}
\mathop{\mathlarger{\int}_{\! 2\,x_N^{}}^\infty} \! \dd z \,
\frac{Z_N^3 Z_\chi \, K_1^{}(z)}
{\big(z^2 - x_\rho^2\big)\rx{1pt}{$^{\!2}$} + x_\rho^2 \, \gamma_\rho^2}
\label{eq:Rate}
~,
\end{eqnarray}
where $n_N^{\rm eq}$ is the equilibrium number density, $\langle \sigma v_{\rm rel} \rangle$ is the thermally averaged annihilation cross section and $N_{k\rm M} \equiv N_k + N_k^c$ denotes the four-component Majorana HNL.\,\,In the second line of Eq.\,\eqref{eq:Rate}, $\lambda_\phi$ is the quartic coupling of $\phi$, $Z_a^{} = (z^2 - 4\,x_a^2)^{1/2}$ with $x_a^{} = m_a^{}/T$, $\gamma_\rho^{} = \Gamma_\rho^{}/T$ with $\Gamma_\rho^{}$ being the total decay rate of $\rho$, and $K_1^{}(z)$ is the modified Bessel function of the second kind of order 1.

To solve Eq.\,\eqref{Boltzmann}, we introduce the comoving yield $Y_\chi \equiv n_\chi/s$, where $s$ is the entropy density.\,\,Using $d\,T/d\,t \simeq - \, {\cal H}\, T$ and $x \equiv m_\chi/T$, we obtain
\begin{eqnarray}
\frac{d \, Y_\chi}{d \, x}
\,=\,
\frac{135 \sqrt{10}  \,\, m_\tx{Pl}^{} \, x^4 \, {\cal R}_{N \! N \to \chi \, \chi}^{}(x)}
{2\,\pi^3 g_{\ast s}^{}(x) \sqrt{g_\ast^{}(x)} \,\, m_\chi^5}
\label{eq:dYdx}
~,
\end{eqnarray}
where $g_\ast^{}$ and $g_{\ast s}^{}$ denote the effective energy and entropy degrees of freedom, respectively, and $m_{\rm Pl}^{} \simeq 2.4 \times 10^{18}\,\text{GeV}$ is the reduced Planck mass.\,\,The present yield $Y_\chi(x_\infty)$ is obtained by integrating Eq.\,\eqref{eq:dYdx} with the initial condition $Y_\chi(T_R^{})=0$, as appropriate for freeze-in production.\,\,The evolution of the comoving DM yield and the reaction rate is shown in Fig.\,\ref{fig:Ychivsx} for a benchmark point with $m_\chi^{} = 1\,\,\text{GeV}$, $m_N^{} = 1\,\,\text{TeV}$, $m_\rho^{} = 10\,\,\text{TeV}$, and $\vphi^{} = 8 \times 10^{12}\,\,\text{GeV}$.\,\,This benchmark point reproduces the DM relic abundance $\Omega_{\rm DM}^\tx{obs} \hat{h}^2 \simeq 0.12$ observed by the Planck Collaboration~\cite{Planck:2018vyg}.\,\,As shown, the DM yield freezes in once the Hubble expansion rate exceeds the production rate.

In the parameter region of interest, an analytic estimate can be obtained.\,\,In particular, for $\Gamma_\rho \ll m_\rho$, the narrow-width approximation applies, and the reaction rate in Eq.\,\eqref{eq:Rate} can be approximately given by 
\begin{eqnarray}
{\cal R}_{N \! N \to \chi \, \chi}^{}
\,\simeq\,
\frac{m_\rho^3 \, m_N^2 \,T}{16 \,\pi^3 \vphi^2}
K_1^{}(x_\rho) \,
{\cal B}_{\rho \to \chi \chi}
\label{eq:RateApp}
~,
\end{eqnarray}
where ${\cal B}_{\rho \to \chi \chi} \equiv \Gamma_{\rho \to \chi \chi}/\Gamma_\rho \simeq 1$.\,\,Substituting it into Eq.\,\eqref{eq:dYdx} and integrating, we obtain
\begin{eqnarray}
Y_\chi(x_\infty^{})
\,\simeq\, 
\frac{405 \sqrt{10}}{(2\,\pi)^5}
\frac{m_N^2 \, m_\tx{Pl}^{}}
{g_{\ast s}^{} \sqrt{g_\ast^{}} \,\, \vphi^2 \, m_\rho^{}}
~,
\label{eq:Ychi}
\end{eqnarray}
which is independent of $m_\chi$. 

Matching to the observed value $Y_\tx{DM}^{} \simeq 4.4 \times 10^{-10} (m_\tx{DM}^{}/\tx{GeV})^{-1}(\Omega_\tx{DM}^\tx{obs} \hat{h}^2/0.12)$~\cite{Planck:2018vyg}, we find
\begin{eqnarray}
\hs{-0.5cm}
\Omega_\chi \hat{h}^2
&\simeq&
0.12 
\bigg(\frac{m_\chi}{1\,\,\tx{GeV}}\bigg)
\bigg(\frac{m_N^{}}{10^3\,\,\tx{GeV}}\bigg)^{\hs{-0.12cm} 2}
\nn
&&\times
\bigg(\frac{m_\rho}{2 \times 10^4\,\,\tx{GeV}}\bigg)^{\hs{-0.12cm} -1}
\bigg(\frac{\vphi^{}}{6 \times 10^{12}\,\,\tx{GeV}}\bigg)^{\hs{-0.12cm} -2}
,
\end{eqnarray}
in agreement with numerical results.\,\,We have also verified consistency with \texttt{micrOMEGAs}~\cite{Belanger:2018ccd}.

\begin{figure}[t!]
\begin{center}
\includegraphics[width=0.48\textwidth]{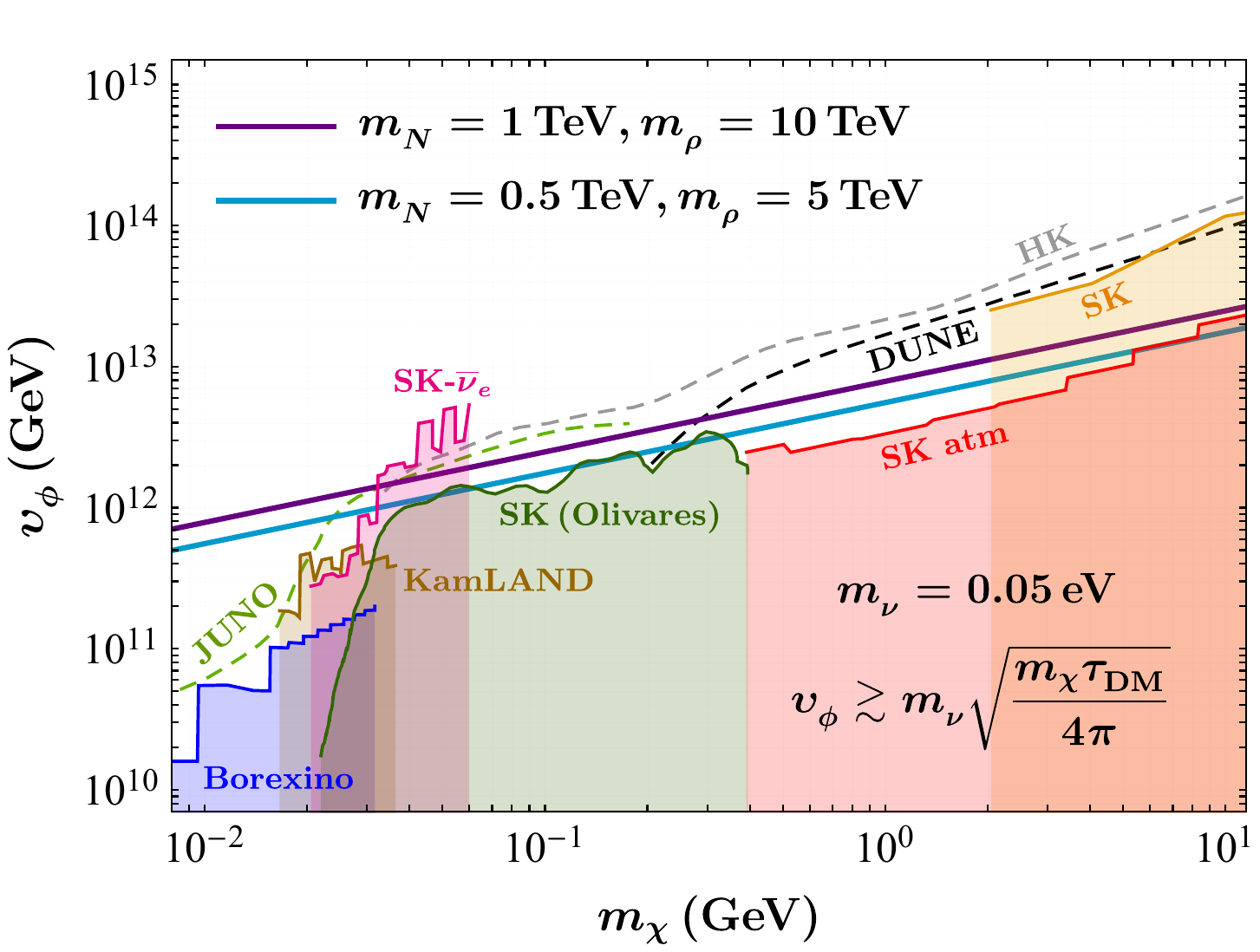}
\vs{-0.5cm}
\caption{Allowed parameter space in the $(m_\chi, \vphi^{})$ plane, where the shadowed regions are excluded by up-to-date observations, the dashed curves represent the future sensitivities, and the oblique lines are benchmark points that satisfy the DM relic density.}
\label{fig:vPhivsmchi}
\end{center}
\end{figure}

As discussed above, conventional DM direct searches are largely insensitive in this model due to the exceptionally large symmetry-breaking scale, $\vphi^{}$, which results in a suppression of the coupling between DM and SM particles.\,\,Nevertheless, a distinctive and potentially observable signal arises from the decay of DM into active neutrinos.\,\,The decay rate of a DM particle into a pair of Majorana neutrinos is given by~\cite{Mohapatra:2021ozu}
\begin{eqnarray}
\Gamma_{\chi \,\to\, \nu_\tx{M}^{} \nu_\tx{M}^{}}
\,=\,
\frac{m_\nu^2}{4\,\pi\,\vphi^2} \, m_\chi
\label{eq:Gammachi}
~,
\end{eqnarray}
where $\nu_\tx{M}^{} = \nu + \nu^\tx{c}$ is a four-component Majorana neutrino.\,\,The decay width is exceedingly suppressed by the combination of the tiny neutrino mass and the large scale $\vphi^{}$, guaranteeing the cosmological stability of the pNGB DM on timescales far exceeding the age of the Universe.

For illustration, the DM lifetime, $\tau_\chi \simeq \Gamma_{\chi \to \nu_\tx{M}^{} \nu_\tx{M}^{}}^{-1}$ (in the limit $m_N^{} \gg m_\chi$), can be expressed as
\begin{eqnarray}
\tau_\chi^{}
\,\simeq\,
10^{23}\,\,\tx{sec}
\bigg(\frac{m_\nu^{}}{0.1\,\,\tx{eV}}\bigg)^{\hs{-0.12cm} 2}
\bigg(\frac{m_\chi}{1\,\,\tx{GeV}}\bigg)
\bigg(\frac{\vphi^{}}{10^{13}\,\,\tx{GeV}}\bigg)^{\hs{-0.12cm} -2}
~,
\label{eq:DMlifetime}
\end{eqnarray}
which satisfies the experimental lower bound on DM lifetime from Super-Kamiokande (SK)~\cite{Super-Kamiokande:2015qek}.\,\,This extreme longevity highlights the challenge of detecting such DM through conventional astrophysical or laboratory searches, but it also points to a clean and well-defined target for neutrino observatories seeking extremely rare signals.

In Fig.\,\ref{fig:vPhivsmchi}, we translate Eq.\,\eqref{eq:Gammachi} into a lower bound on the symmetry-breaking scale, $\vphi^{}$, as a function of the DM mass, $m_\chi$, based on the analysis in Ref.\,\cite{Arguelles:2022nbl}.\,\,The shaded region corresponds to the excluded region from various experiments, while the dashed curves represent future sensitivities.\,\,This mapping provides a direct connection between the pNGB DM parameter space and experimental observables, illustrating that even in scenarios with ultra-heavy $\vphi^{}$, neutrino decay channels remain a promising probe of otherwise elusive DM.

\textit{Results.}\textemdash We numerically solve Eq.\,\eqref{eq:dYdx} and fix the resulting relic abundance to $\Omega_\chi \hat{h}^2 = 0.12$.\,\,The required symmetry-breaking scale, $\vphi$, is then determined as a function of the DM mass $m_\chi$ with fixing $m_N^{} = 1\,\,\tx{TeV}$ and $m_\rho^{} = 10\,\,\tx{TeV}$ ($m_N^{} = 0.5\,\,\tx{TeV}$ and $m_\rho^{} = 5\,\,\tx{TeV}$), as shown by solid purple (cyan) line in Fig.\,\ref{fig:vPhivsmchi}. 

For TeV-scale HNLs, we find that the DM mass $m_\chi^{} \gtrsim 2\,\,\tx{GeV}$ are excluded by Super-Kamiokande (SK) searches~\cite{Frankiewicz:2017trk}, while the intermediate range $30\,\,\text{MeV} \lesssim m_\chi \lesssim 60\,\,\text{MeV}$ is disfavored by constraints on the $\bar{\nu}_e$ flux~\cite{Linyan:2018rmj}.\,\,In contrast, the sub-GeV region remains consistent with current neutrino data~\cite{Super-Kamiokande:2015qek,Olivares-DelCampo:2017feq}, and interestingly lies in the region that can be probed by ongoing experiments such as JUNO~\cite{Akita:2022lit} and upcoming neutrino facilities such as Hyper-Kamiokande~\cite{Bell:2020rkw} and DUNE~\cite{Arguelles:2019ouk}.\,\,This behavior reflects the interplay between the freeze-in production rate and late-time decay signatures, which become increasingly constrained at higher masses.

Remarkably, the same TeV-scale HNLs that govern DM production and decay also generate light neutrino masses, linking cosmology to the seesaw mechanism.\,\,These states can be directly probed at high-energy colliders through their mixing with SM leptons~\cite{Atre:2009rg, Das:2012ze, Bolton:2019pcu}.\,\,At the LHC, HNLs are produced via charged-current processes such as $q\bar{q}' \to W^\ast \to N\ell$, followed by decays $N \to \ell \, W,\; \nu Z,\; \nu h$~\cite{Atre:2009rg,delaTorre:2024urj}.\,\,For Majorana HNLs, this leads to the characteristic same-sign dilepton plus jets signature with suppressed missing energy, providing a smoking-gun signal of the lepton-number violation~\cite{Keung:1983uu}.\,\,Current LHC searches have already placed constraints on the active–sterile mixing down to $|V_{\ell N}^{}|^2 \simeq \xi^2 \sim 10^{-1}$ for $m_N^{} \sim \mathcal{O}(\mathrm{TeV})$~\cite{CMS:2018iaf,CMS:2024xdq}.\,\,Looking ahead, the HL-LHC with an integrated luminosity of $3\,\,\mathrm{ab}^{-1}$ is expected to improve the sensitivity to $|V_{\ell N}^{}|^2 \sim 10^{-2}$ in this mass range~\cite{Pascoli:2018heg}.\,\,Future multi-TeV lepton colliders provide complementary and significantly cleaner probes.\,\,Processes such as $e^+e^- \to N \nu$, $\mu^+\mu^- \to N\nu$, or $\mu^+ e^- \to N \nu$ benefit from reduced backgrounds and precise kinematic reconstruction~\cite{Banerjee:2015gca,Das:2018usr,Chakraborty:2018khw,Mekala:2022cmm,
Mekala:2023diu,Das:2024kyk}.\,\,In particular, a 3-TeV muon collider with an integrated luminosity 
of $1\,\,\mathrm{ab}^{-1}$ can reach sensitivities down to $|V_{\ell N}^{}|^2 \sim 10^{-6}$ for TeV-scale HNLs~\cite{Mekala:2023diu}. 

\textit{Conclusion and Outlook.}\textemdash In this Letter, we have demonstrated that a minimal extension of the inverse seesaw model can simultaneously account for neutrino mass generation and DM phenomenology in a tightly correlated and predictive framework.\,\,The spontaneous breaking of lepton number gives rise to a pseudo-Nambu-Goldstone boson that serves as a DM candidate, while a small explicit breaking term generates its mass.\,\,Remarkably, the same TeV-scale HNLs responsible for neutrino mass produce DM via freeze-in and mediate its decay into neutrinos.

This structure leads to a direct connection among neutrino masses, DM relic abundance, and DM lifetime.\,\,The \UL symmetry breaking scale $\vphi^{}$ controls the DM decay, ensuring its cosmological stability while allowing potentially observable neutrino signals.\,\,Requiring the observed DM relic density and current bounds on its lifetime significantly constrains the parameter space, pointing toward sub-GeV DM when HNLs lie at the TeV scale.

An important implication of this framework is its multi-front testability.\,\,HNLs in the TeV range can be probed in present and future collider experiments, while DM decay into neutrinos can be searched for at next-generation neutrino detectors.\,\,A discovery of TeV-scale HNLs would therefore imply a correlated prediction for a neutrino signal from DM decay, providing a concrete experimental pathway to test the common origin of neutrino mass and DM.

Finally, we comment on several possible extensions of this model.\,\,The TeV-scale HNLs introduced here may also account for the baryon asymmetry of the Universe via resonant leptogenesis in a quasi-degenerate regime, where CP-violating decays can generate an asymmetry with suppressed washout despite prior thermalization~\cite{Pilaftsis:1997jf,Pilaftsis:2003gt}.\,\,If the symmetry breaking associated with $\vphi$ proceeds through a first-order phase transition, the resulting bubble nucleation, expansion, and collision can source a stochastic gravitational-wave background potentially observable at future detectors~\cite{Witten:1984rs,Hogan:1986qda}.\,\,Alternatively, the feeble interactions of the HNLs may induce a stochastic gravitational-wave background from thermal fluctuations in the early plasma~\cite{Drewes:2023oxg}.\,\,These complementary signatures highlight the potential of this scenario to be probed across cosmological, astrophysical, and collider frontiers.

\textit{Acknowledgments.}\textemdash We are grateful to Koji Tsumura, Jongkuk Kim, Riasat Sheikh and Tzu-Chiang Yuan for the useful discussions.\,\,This work was partially supported by the National Science and Technology Council under Grant No.~NSTC114-2112-M-002-020-MY3 (CWC), the Ministry of Education (Higher Education Sprout Project NTU-114L104022-1), the National Center for Theoretical Sciences of Taiwan, and the Vietnam National Foundation for Science and Technology Development (NAFOSTED) under Grant No.~103.01-2023.50 (VQT).

\clearpage
\appendix 
\onecolumngrid
\begin{center}
\bf \large Supplemental Material
\end{center}

\section{Scalar potential}

The scalar potential invariant under the \UL symmetry is given by
\begin{eqnarray}
{\cal V}_{\rm inv}
\,=\,
- \frac{1}{2} \mu_h^2 |H|^2
- \frac{1}{2} \mu_\phi^2 |\phi|^2
+ \frac{1}{2} \lambda_h^{} |H|^4
+ \frac{1}{2} \lambda_\phi^{} |\phi|^4
+ \lambda_{h \phi}^{} |H|^2 |\phi|^2
\label{eq:potential}
~,
\end{eqnarray}
where $\mu_h^2,\,\mu_\phi^2,\,\lambda_h^{},\,\lambda_\phi^{},\,\lambda_{h\phi}^{} > 0$.\,\,Hermiticity of ${\cal V}_{\rm inv}$ requires all the parameters to be real.

With these choices, both $H$ and $\phi$ acquire vacuum expectation values (VEVs).\,\,In particular, $\langle \phi \rangle$ spontaneously breaks the global \UL symmetry in the early Universe.\,\,According to Goldstone's theorem~\cite{Goldstone:1961eq}, this results in a massless Nambu-Goldstone boson (NGB), identified with the phase of $\phi$.\,\,The radial component of $\phi$, being massive, typically decays rapidly into NGBs or SM particles and therefore does not constitute a viable dark matter (DM) candidate.

Instead, one can introduce a linear \UL soft-breaking term, as given in Eq.\,(5) of the main text, which generates a mass for the pNGB and makes it a viable DM candidate.\,\,Alternatively, a quadratic \UL soft-breaking term, $\phi^2$, can be considered~\cite{Gross:2017dan}.\,\,In this case, the DM-nucleon scattering amplitude is suppressed at tree level in the zero momentum-transfer limit, allowing the model to evade current direct detection bounds~\cite{Cai:2021evx,Coito:2021fgo}. 
However, such a scenario may suffer from the cosmic domain wall problem~\cite{Zeldovich:1974uw}.

In the limit $\lambda_{h\phi}^{} \to 0$, the SM Higgs boson decouples from the heavier CP-even scalar $\rho$. 
The scalar masses are then given by 
\begin{eqnarray}
m_h^{2} 
\,=\, \lambda_h^{} \vh^2 
\,\simeq\, 
(125\,\,\tx{GeV})^2
~,\quad
m_\rho^2
\,=\,
\lambda_\phi \vphi^2 + m_\chi^2
~,~ {\text{and} } ~~
m_\chi^2
\,=\,
\frac{\kappa_\phi^3}{2\,\vphi}
~.
\end{eqnarray}
The self-interactions of $\rho$ and $\chi$ follow from
\begin{eqnarray}
{\cal V}_{\rm inv}
\,\supset\,
\frac{1}{2} \lambda_\phi^{} \, \vphi^{} \, \rho^3
+
\frac{1}{2} \lambda_\phi^{} \, \vphi^{} \, \rho \, \chi^2
+
\frac{1}{4} \lambda_\phi^{} \, \rho^2 \chi^2
\label{eq:sint}
~,
\end{eqnarray}
which respects an accidental $\mathbb{Z}_2$ symmetry under which $\rho \to \rho$ and $\chi \to -\chi$.

\section{Neutrino masses in the inverse seesaw and Yukawa interactions}

We derive the masses and interactions of the active neutrinos and heavy neutral leptons (HNLs) after electroweak and \UL symmetry breaking.\,\,For simplicity, we consider a single fermion generation, such that the Yukawa couplings and Majorana mass reduce to complex parameters.\,\,Substituting the scalar VEVs into Eq.\,(1) of the main text yields
\begin{eqnarray}
{\cal L}
\,\supset\,
- \,\, m_D^{} \, \overline{\hat{\nu}_L^{}} \, N_R^{}
- m_N^{} \, \overline{S_L^{}} \, N_R^{}
- \frac{1}{2} \mu_S^{} \, \overline{S_L^{}} \, S_L^c
+ \text{H.c.}
\label{eq:ISS}
~,
\end{eqnarray}
where $m_D^{} = {\cal Y}_D^{} \vh^{}/\sqrt{2}$ and $m_N^{} = {\cal Y}_N^{} \vphi^{}/\sqrt{2}$.
In the flavor basis $\Psi_f = \big(\hat{\nu}_L^{}~ N_R^c~ S_L^{}\big){}^\tx{\!T}$, the Lagrangian can be written as
\begin{eqnarray}
{\cal L}
\,\supset\,
- \, \frac{1}{2} \, \overline{\Psi_f^c} \, {\cal M}_f \, \Psi_f
+ \text{H.c.}
\end{eqnarray}
with the mass matrix
\begin{eqnarray}
{\cal M}_f
\,=\,
\begin{pmatrix}
0 & m_D^{} & 0
\\[0.1cm]
m_D^{} & 0 & m_N^{}
\\[0.1cm]
0 & m_N^{} & \mu_S^{}
\end{pmatrix}
~,
\end{eqnarray}
where phases have been absorbed into the fermion fields such that all entries are real and positive.

Diagonalizing ${\cal M}_f$ yields the mass eigenstates $\Psi_m = \big(\nu \,~ N_1~ N_2\big){}^\tx{\!T}$, related to the flavor basis by $\Psi_m = \,{\cal U}^\dagger \Psi_f$, where ${\cal U}$ is a unitary matrix.\,\,To account for the observed small neutrino masses, we adopt the hierarchy $\mu_S^{} \ll m_D^{} \ll m_N^{}$.\,\,Expanding to the leading order in $\xi \equiv m_D^{}/m_N^{}$ and $\omega \equiv \mu_S^{}/m_N^{}$, the mixing matrix is
\begin{eqnarray}
{\cal U}
\,\simeq\,
\begin{pmatrix}
\,\, 1 - \xi^2/2 & \xi/\sqrt{2} & \xi/\sqrt{2} ~~
\\[0.1cm]
\xi \, \omega & -1/\sqrt{2} & 1/\sqrt{2}  ~~
\\[0.1cm]
- \,\xi & 1/\sqrt{2} & 1/\sqrt{2}  ~~
\end{pmatrix}
{\cal D}
~,
\label{eq:Uni}
\end{eqnarray}
where ${\cal D} = \mathrm{diag}(1, i, 1)$ ensures positive mass eigenvalues~\cite{Kayser:2002qs}, such that ${\cal U}^\tx{T} \! {\cal M}_f \,\, {\cal U} = \mathrm{diag}(m_{\nu}^{},\,m_{N_1}^{},\, m_{N_2}^{})$.\,\,The mass eigenvalues are then given by
\begin{eqnarray}
m_{\nu}^{}
\,\simeq\,
\mu_S^{} \,\frac{m_D^2}{m_N^2}
~,\quad
m_{N_{1,2}}^{}
\,\simeq\,
m_N^{} + \frac{m_D^2}{2 \, m_N^{}} \mp \frac{\mu_S^{}}{2}
~,
\label{eq:M123}
\end{eqnarray}
demonstrating that the light neutrino mass is suppressed by both the small lepton-number-violating parameter $\mu_S^{}$ and the ratio $m_D^{}/m_N^{}$, while the heavy states form a quasi-Dirac pair with a small mass splitting controlled by $\mu_S^{}$.

Lastly, we derive the scalar couplings to the fermions.\,\,From Eq.\,(1), one can easily obtain 
\begin{eqnarray}
{\cal L}
\,\supset\,
- \,\frac{m_N^{}}{\vphi^{}} \, \overline{S_L^{}} \,N_R^{} \, \big(\,\rho + i \, \chi \big)
+ \tx{H.c.}
\label{eq:SF}
\end{eqnarray}
Next, using $\Psi_{\!f} = \,{\cal U} \,\Psi_{\! m}^{}\!\!$ with the form in Eq.\,\eqref{eq:Uni}, we have
\begin{eqnarray}
N_R^{}
\Eq
\xi \, \omega \, \nu^c
+
\frac{i}{\sqrt2} \, N_1^c
+
\frac{1}{\sqrt2} \, N_2^c
~,\quad
\\
S_L^{}
\Eq
- \, \xi  \, \nu
+
\frac{i}{\sqrt2} \, N_1^{}
+
\frac{1}{\sqrt2} \, N_2^{}
~.
\end{eqnarray}
Plugging them into Eq.\,\eqref{eq:SF} with $m_\nu^{} \! \simeq m_N^{} \, \omega \, \xi^2$, the  interaction of $\chi$ with the active neutrino is given by
\begin{eqnarray}
{\cal L}
\,\supset\,
i\,\frac{m_\nu^{}}{\vphi^{}}  \,
\overline{\nu_\tx{M}^{}}  \, \gamma^5 \nu_\tx{M}^{} \, \chi
\label{eq:nuchi}
~,
\end{eqnarray}
where $\nu_\tx{M}^{} = \nu + \nu^\tx{c}$ is a four-component Majorana neutrino.\,\,Likewise, the interactions between the scalars and HNLs are extracted as
\begin{eqnarray}
{\cal L}
\,\supset\,
- \, \frac{m_N^{}}{2 \, \vphi^{}} 
\sum_{k = 1,2}
\overline{N_{k\tx{M}}^{}} \, \big(\,\rho + i \, \gamma^5 \chi\big) N_{k\tx{M}}^{}
\label{eq:Nrhochi}
~,
\end{eqnarray}
where $N_{k\tx{M}}^{} = N_k^{} + N_k^\tx{c}$ are heavy four-component Majorana fermions with masses $\sim m_N^{}$.

\section{Thermal production of HNL}

\begin{figure}[t!]
\begin{center}
\includegraphics[width=0.48\textwidth]{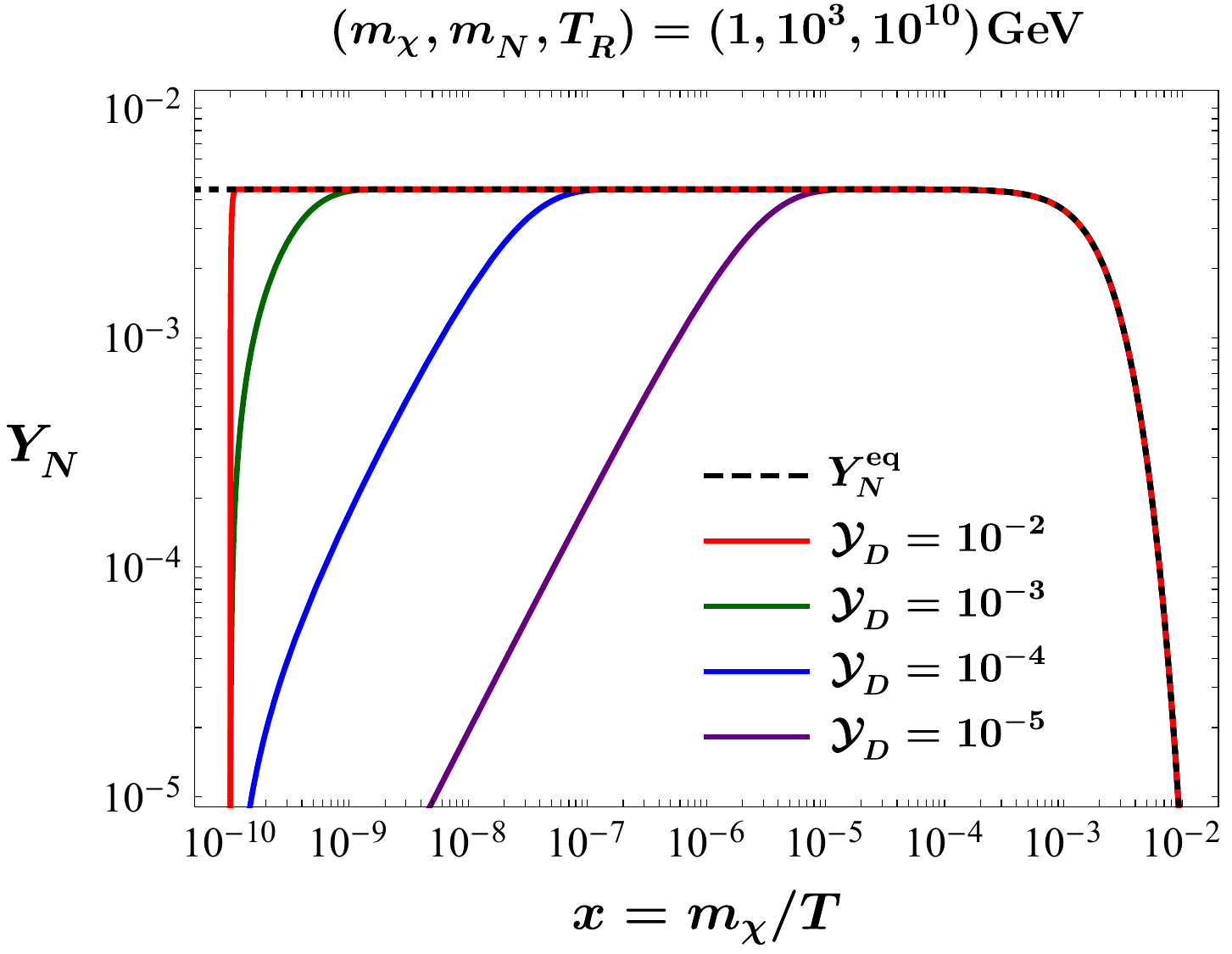}
\caption{Time evolution of the comoving number density of the HNL.}
\label{fig:YN}
\end{center}
\end{figure}

In this model, the HNLs can be thermally produced from the SM plasma at the reheating temperature via the Yukawa coupling ${\cal Y}_D^{}$ given in Eq.\,(1) of the main text.\,\,One of the production channels is the decay of the SM Higgs boson, $H \to N E$.\,\,The Boltzmann equation for the number density of the HNL is given by
\begin{eqnarray}
\frac{d \, n_N^{}}{d \, t} + 3 \, {\cal H} \, n_N^{}
\,=\,
n^\tx{eq}_H \big\langle \Gamma_{H \to N L}^{} \big\rangle
\bigg(1 - \frac{n_N^{}}{n_N^\tx{eq}}\bigg)
+ 
\cdots
~,
\end{eqnarray}
where $\cdots$ denotes other production channels, and $\big\langle \Gamma_{H \to N L}^{}\big\rangle$ is the thermally averaged decay rate
\begin{eqnarray}
\big\langle \Gamma_{H \to N L}^{} \big\rangle
\,=\,
\frac{K_1^{}\big[m_H^{}(T)/T\big]}{K_2^{}\big[m_H^{}(T)/T\big]}\,
\Gamma_{H \to N L}^{}
\end{eqnarray}
with the non-thermally averaged  decay rate (assume $m_H^{}(T) \gg m_{N,L}^{}$)
\begin{eqnarray}
\Gamma_{H \to N L}^{}
\,=\,
\frac{{\cal Y}_D^2 \, m_H^{}(T)}{8\,\pi}
~.
\end{eqnarray}
Note that $m_H^{}(T)$ here is the thermal mass of the SM Higgs boson since the HNLs were produced at temperatures above the energy scale of the electroweak symmetry breaking~\cite{Abe:2020ldj}.\,\,In terms of the comoving yield $Y_N^{} = n_N^{}/s$ and $x$, the corresponding Boltzmann equation becomes
\begin{eqnarray}
\frac{d \, Y_N^{}}{d \, x}
\,=\,
\frac{Y_H^\tx{eq} \big\langle \Gamma_{H \to N L}^{} \big\rangle}{x \, {\cal H}}
\bigg(1 - \frac{Y_N^{}}{Y_N^\tx{eq}}\bigg)
+ 
\cdots
\label{eq:BoltzmannN}
~,
\end{eqnarray}
where $Y^\tx{eq}_a = n^\tx{eq}_a/s$.\,\,Assuming a negligible initial abundance of $N$, we solve Eq.\,\eqref{eq:BoltzmannN} with a benchmark point $m_\chi =1\,\,\tx{GeV},\,\,m_N^{} = 1\,\,\tx{TeV}$, and $T_R^{} =10^{10}\,\,\tx{GeV}$, and show the result in Fig.\,\ref{fig:YN}.\,\,Compared with Fig.\,2 in the main text, our numerical calculation is justified if ${\cal Y}_D^{} \gtrsim 10^{-5}$ and $T_R^{} \gg m_N^{}$.\,\,Note that the lower bound of ${\cal Y}_D^{}$ can be much smaller if more production processes are included.\,\,However, we have to choose a sufficiently large value of ${\cal Y}_D^{}$ since the neutrino mass formula in the main text is valid when $\mu_S^{} \ll m_D^{} \sim {\cal Y}_D^{} \vh^{}$.\,\,On the other hand, electroweak precision data require $\xi^2 = \big(m_D^{} / m_N^{})^2 \lesssim 2.5 \times 10^{-3}$ for $m_N^{} \gtrsim 1\,\,\tx{GeV}$, which then places an upper bound on ${\cal Y}_D^{}$~\cite{Bolton:2019pcu}.\,\,With these two restrictions, a proper benchmark point for the neutrino mass generation is given as follows
\begin{eqnarray}
m_\nu^{} 
\,\simeq\,
0.1\,\,\text{eV}
\bigg(\frac{\mu_S^{}}{0.1\,\,\text{MeV}}\bigg)
\bigg(\frac{\xi^{}}{10^{-3}}\bigg)^{\hs{-0.12cm}2}
~,
\end{eqnarray}
where
\begin{eqnarray}
\xi
\,\simeq\,
10^{-3}
\bigg(\frac{{\cal Y}_D^{}}{6 \times 10^{-3}}\bigg)
\bigg(\frac{m_N^{}}{1\,\,\text{TeV}}\bigg)^{\hs{-0.12cm}-1}
~,
\end{eqnarray}
which can be probed by future muon colliders~\cite{Mekala:2023diu}, with
${\cal Y}_D^{} \sim 6 \times 10^{-3} \big(m_D^{}/\tx{GeV}\big)$.

\vs{0.3cm}

\twocolumngrid

\bibliography{refs}

\end{document}